\documentclass[times,twocolumn,final]{elsarticle}

\usepackage{lineno,hyperref}
\usepackage{multirow}
\usepackage[dvipsnames]{xcolor}
\modulolinenumbers[5]

\journal{Medical Image Analysis}

\biboptions{authoryear}

\begin{document}

\begin{frontmatter}

\title{Large-scale Gastric Cancer Screening and Localization\\Using Multi-task Deep Neural Network}

% \author{author1\fnref{myfootnote}}
% \author{Hong Yu, Xiaofan Zhang, Lingjun Song, Liren Jiang, Xiaodi Huang, Wen Chen, Chenbin Zhang, Jiahui Li, Jiji Yang, Zhiqiang Hu, Qi Duan, Wanyuan Chen, Xianglei He, Jinshuang Fan, Weihai Jiang, Li Zhang, Chengmin Qiu, Minmin Gu, Weiwei Sun, Yangqiong Zhang, Guangyin Peng, Weiwei Shen, Guohui Fu}

% \address{address1 29, country}

% \tnotetext[mytitlenote]{Equal Contribution.}
% \tnotetext[myfootnote]{Equal Contribution.}
% \fntext[myfootnote]{Since 1880.}

% \author[mymainaddress,mysecondaryaddress]{Elsevier Inc}

\author[my1address]{Hong Yu\corref{myfirstauthor}}

\author[my2address]{Xiaofan Zhang\corref{myfirstauthor}}

\author[my1address]{Lingjun Song\corref{myfirstauthor}}
\author[my1address]{Liren Jiang\corref{myfirstauthor}}

\author[my2address]{Xiaodi Huang\corref{myfirstauthor}}

\author[my2address]{Wen Chen}
\author[my2address]{Chenbin Zhang}
\author[my2address]{Jiahui Li}

\author[my1address]{Jiji Yang}

\author[my2address]{Zhiqiang Hu}
\author[my2address]{Qi Duan}

\author[my3address]{Wanyuan Chen}
\author[my3address]{Xianglei He}

\author[my4address]{Jinshuang Fan}
\author[my4address]{Weihai Jiang}

\author[my5address]{Li Zhang}
\author[my5address]{Chengmin Qiu}
\author[my5address]{Minmin Gu}
\author[my5address]{Weiwei Sun}
\author[my5address]{Yangqiong Zhang}

\author[my6address]{Guangyin Peng}

\author[my1address]{Weiwei Shen\corref{mycorrespondingauthor}}
\author[my1address]{Guohui Fu\corref{mycorrespondingauthor}}

\cortext[myfirstauthor]{Equal contribution}
\cortext[mycorrespondingauthor]{Corresponding author}
\cortext[myfunding]{This work is a part of the key projects in Shanghai Science \& Technology Pillar Program for Biomedicine (NO14431904700) and supported by the funding of Science and Technology Commission Shanghai Municipality(19511121400).}
\ead{guohuifu@shsmu.edu.cn}

\address[my1address]{Pathology Center, Shanghai General Hospital/Faculty of Basic Medicine; Key Laboratory of Cell Differentiation and Apoptosis of Chinese Ministry of Education, Institutes of Medical Sciences, Shanghai Jiao Tong University School of Medicine, Shanghai, China}
\address[my2address]{SenseTime Research}
\address[my3address]{Department of Pathology, Zhejiang Provincial People’s Hospital, People’s Hospital of Hangzhou Medical College, Zhejiang, China}
\address[my4address]{Department of Pathology, Sijing Hospital of Shanghai Songjiang District, Shanghai, China}
\address[my5address]{Department of Pathology, Songjiang District Central Hospital, Shanghai, China}
\address[my6address]{Department of Pathology, People’s Hospital of Dong Tai, Jiang Su Province, China}

\begin{abstract}
Gastric cancer is one of the most common cancers, which ranks third among the leading causes of cancer death. 
Biopsy of gastric mucosa is a standard procedure in gastric cancer screening test. 
However, manual pathological inspection is labor-intensive and time-consuming. Besides, it is challenging for an automated algorithm to locate the small lesion regions in the gigapixel whole-slide image and make the decision correctly.
To tackle these issues, we collected large-scale whole-slide image dataset with detailed lesion region annotation and designed a whole-slide image analyzing framework consisting of 3 networks which could not only determine the screening result but also present the suspicious areas to the pathologist for reference. 
Experiments demonstrated that our proposed framework achieves sensitivity of $97.05\%$ and specificity of $92.72\%$ in screening task and Dice coefficient of $0.8331$ in segmentation task. Furthermore, we tested our best model in real-world scenario on $10,315$ whole-slide images collected from $4$ medical centers. 
%(XXX10316 original)
\end{abstract}


\end{frontmatter}

% \linenumbers

\section{Introduction}

Gastric cancer remains important cancer worldwide and is responsible for over $1,000,000$ new cases in 2018 and an estimated $783,000$ deaths (equating to 1 in every 12 deaths globally), making it the fifth most frequently diagnosed cancer and the third leading cause of cancer death.
Biopsy of the gastric mucosa is one of the most effective methods of early detection of gastric cancer~\cite{bray2018global}.
% Gastric cancer incidence rates are markedly high in Eastern Asia. 
It is estimated that there are hundreds of millions of gastric biopsy slides need to be examined in China each year, while the number of certified pathologists is only about $10$ thousand, which causes excessive workloads on these pathologists.

To reduce the workload of the pathologists, there are extensive studies focus on pathology image analysis~\cite{zhang2014towards,xu2015stacked,fakhry2016residual,li2018large,hu2018unsupervised,duan2020sensecare}.
In recent years, deep neural network techniques have achieved remarkable performance on a wide range of computer vision tasks, such as image classification~\cite{krizhevsky2012imagenet,he2016deep,szegedy2016rethinking}, object detection~\cite{he2017mask,lin2017focal,redmon2018yolov3}, semantic segmentation~\cite{long2015fully,chen2018encoder}, etc. 
These techniques have been applied in automated pathology image analysis in the past few years.
Unlike natural images, digital pathology images, named whole-slide images (WSIs), are extremely large whose width and height often exceed $100,000$ pixels.  
On the other hand, histological diagnosis requires high accuracy since it is commonly considered  as the gold standard.
As a result,
some of the studies focus on the selected regions of interest (ROIs)~\cite{su2015region,zhang2015high,zhang2015fusing,albarqouni2016aggnet,chen2017dcan,Li2019SignetRC,li2019accurate}, while there are several attempts on analyzing WSI~\cite{barker2016automated,coudray2018classification,lin2018scannet,zhang2019pathologist}.

Besides the difficulties in applying the deep neural network on gigapixel resolution images, the main challenge in examining the WSIs is that the diagnostic results labeled by the pathologists are usually on the slide level in most of the publicly available datasets, while the lesion regions that draw the pathologists' attention are extremely small compared with the size of the WSI. It is tough to train a deep neural network to locate those regions and make the correct decision only using slide level labels such ``positive/negative''.
Therefore, we collect a large dataset that not only has the slide level annotation but also carries the lesion region annotation and design a framework leveraging the detailed supervised information.

To our best knowledge, there have been no studies on automated pathology image analysis with lesion region annotation for gastric cancer. We propose an automated screening framework that could not only provide the screening results, i.e., positive/negative, but also show the suspicious areas to pathologists for further reference.

%------------------------------------------------------------------------
\section{Related Works}

\subsection{Pathology Image Analysis for Gastric Cancer}
Only a few existing works are focusing on analyzing the pathology image from the biopsy of the gastric mucosa for diagnosis.

\cite{cosatto2013automated} designed a semi-supervised multiple instance learning framework that takes $230 \times 230$ microns at magnification of $20X$ and $460 \times 460$ microns at magnification of $10X$ ROIs segmented from tissue units as input and analyzes their color features.

\cite{oikawa2017pathological} proposed a computerized analysis system which first analyzes color and texture features on the entire H\&E-stained section at low resolution to search suspicious areas for cancer (except signet ring cell which is detected by CNN-based method), and then analyzes contour and pixel features at high resolution on selected area and uses a trained SVM to confirm the initial suspicion.

\cite{li2018deep} proposed GastricNet to automatically identify gastric cancer. The network with different architectures for shallow and deep layers was applied on $224\times224$ patches cropped from  $700$ $2048\times2048$ ROIs with a magnification factor of $20X$. 

\cite{yoshida2018automated} compared the classification results of human pathologists and of the e-Pathologist. The e-Pathologist analyzes high-magnification features that characterized the nuclear morphology and texture as well as low-magnification features that characterized the global H\&E stain distribution within an ROI and the appearance of blood cells and gland formation. The classification was modeled as a multi-instance learning problem and solved by training a multi-layer neural network.

\subsection{Whole-Slide Image Analysis}

As the gold standard for various cancer diagnosis, WSIs analysis related techniques have been well-studied in recent years.

\cite{liu2017detecting} presented a CNN framework to aid breast cancer metastasis detection in lymph nodes. The model is based on Inception architecture with careful image patch sampling and data augmentations. Random forest classifier and feature engineering were used for whole-slide classification procedure.

\cite{hou2016patch} proposed to train a decision fusion model to aggregate patch-level predictions given by patch-level CNNs. In addition, authors formulated a Expectation-Maximization based method that automatically locates discriminative patches robustly by utilizing the spatial relationships of patches. 

\cite{zhu2017wsisa} proposed a survival analysis framework which first extract hundreds of patches by adaptive sampling and then group these images into different clusters. An aggregation model was trained to make patient-level predictions based on cluster-level results.

\cite{mercan2017multi} developed a framework to analyze breast histopathology image. Candidate ROIs were extracted from the logs of pathologists’ image screenings based on different behaviors. Class labels were extracted from the pathology forms and slides were modeled as bags of instances which are represented by the candidate ROIs.

Localizing of lesion regions in WSIs by image segmentation related techniques is also a crucial direction for helping pathologists make the correct decision efficiently.

\cite{qaiser2019fast} proposed a tumor segmentation framework based on the concept of persistent homology profiles which models the atypical characteristics of tumor nuclei to localize the malignant tumor regions.  \cite{dong2018reinforced} proposed a reinforcement learning based framework motivated by the zoom-in operation of a pathologist which learns a policy network to decide whether zooming is required in a given ROI.

\bigskip

Our main contributions:
\begin{itemize}
    \item We collect a large-scale dataset for gastric cancer screening and develop a semi-automated annotation system to help obtain the detailed lesion region annotation.
    \item We take advantage of the region annotation by proposing a multi-task network structure which could provide the classification label (screening result) as well as the segmentation mask (suspicious region) simultaneously.
    \item We design a practical framework consisting of $3$ networks to process the high-resolution WSIs, and employ the deformable convolution operation based on the observation of the characteristics of the pathology images.
\end{itemize}

%------------------------------------------------------------------------

\section{Data Collection}

%All images were deidentified according to Health Insurance Portability and Accountability Act Safe Harbor prior to transfer to study investigators. Ethics review and institutional review board exemption was obtained using Quorum Review IRB.

%https://www.leicabiosystems.com/pathologyleaders/an-introduction-to-specimen-preparation/
% 图A：一张WSI的缩略图，特点是：HE染色的标志性粉紫色，注意左边的切块跟右边的切块明显长得不一样，大多数时候他们来自胃的不同部位的取材

% 图B：B是A里的一个切块（tussue）的放大图，注意到这个切块里面有不规则的纹理和一些黏膜肌

% 图C：C是B中的所有病灶的区域，用红色的overlay表示

% 一张WSI的制作过程要经过：取材，固定（tissue fixation），包埋（tissue processing），切片（sectioning），染色（staining），扫描（scanning）等多个步骤。

% 在胃黏膜WSI的制作过程中，取材通常从胃镜gastroscope的过程中挖疑似恶性的组织出来，在切片的时候将这块小组织切成很多薄片，由于胃黏膜的组织块太小了，有效面积通常较小，因此为了防止遗漏病灶，医生会把切出来的这些薄片都码在玻璃片上，制作成一张slide，随后再扫描为WSI。

\begin{figure}[h]
\centering
\includegraphics[width=0.85\linewidth]{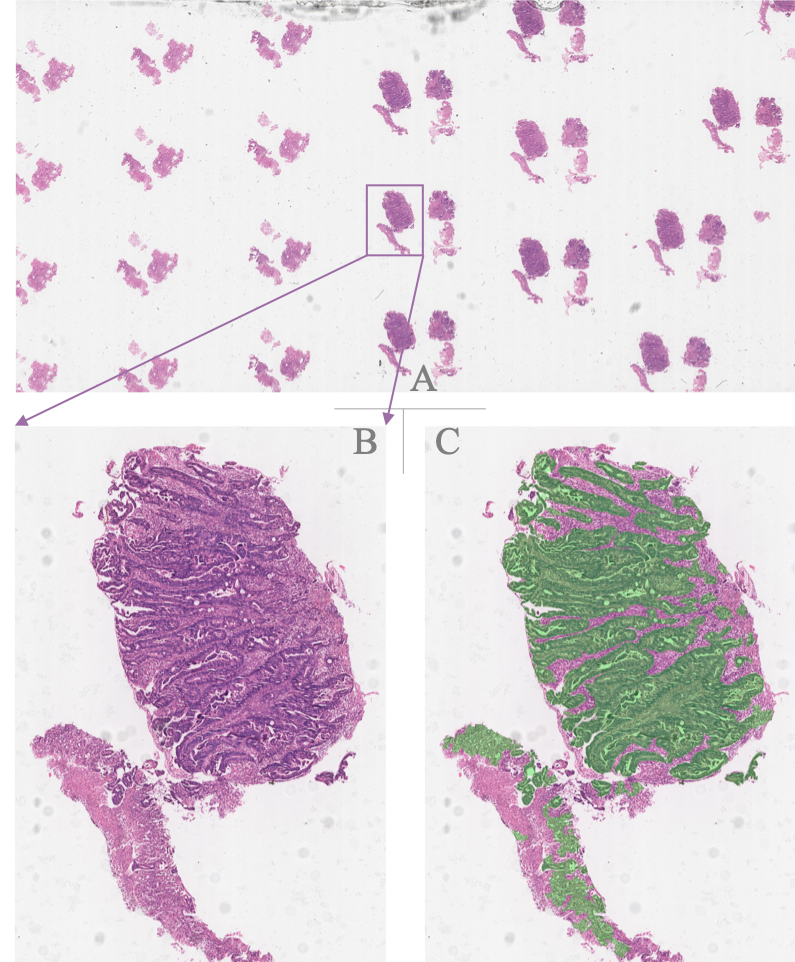}
\vspace{-0.5em}
\caption{A) A typical WSI of gastric biopsy. Pathologists usually cut all the suspicious tissues of gastric mucosa acquired from gastroscope into multiple thin sections and put it on a single glass slide. Hence, sections on the same slide could be the adjacent part of the same tissue or even different tissues. B) One of the enlarged regions of WSI. There are some irregular textures and muscularis mucosae. C) Lesion region annotation. The region considered as malignant is denoted by the green overlay.}
\vspace{-1em}
\label{fig:dataset}
\end{figure}

\begin{figure*}[t]
\centering
\includegraphics[width=1\linewidth]{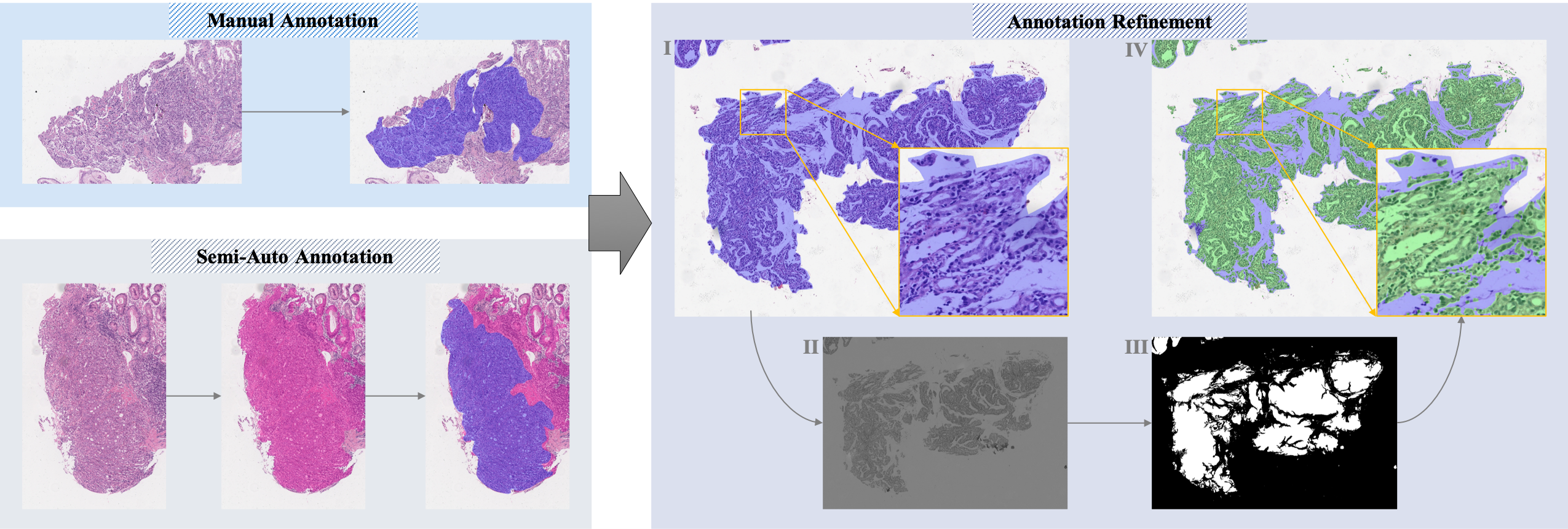}
\vspace{-1em}
\caption{Our annotation pipeline. The pipeline consists of two parts, namely the initial annotation and the refinement. The initial annotation part (on the left) has two routes:  manul annotation (blue overlay) and semi-auto annotation (pink denotes the output of the U-Net, blue denotes the annotation after pathologists' editing). The refinement part (on the right) shows I) the annotation from pathologists (blue), II) output of color deconvolution, III) binary image after applying Otsu thresholding. IV) the final ground truth annotation (green) after refinement.}
\label{fig:seg_anno_auto}
\end{figure*}

\subsection{Data Acquisition}
All the annotated slides are automatically scanned using the digital pathology scanner Leica Aperio AT2 at $20X$ magnification ($0.50 um/pixel$) and labeled by pathologists under the supervision of the experts in Shanghai General Hospital.

The slide-level annotation is either ``positive'' (refers to the malignant sample such as low-grade intraepithelial neoplasia, high-grade intraepithelial neoplasia, adenocarcinoma, signet ring cell carcinoma, and poorly cohesive carcinoma) or ``negative'' (refers to the benign sample such as chronic atrophic gastritis, chronic non-atrophic gastritis, intestinal metaplasia, gastric polyps, gastric mucosal erosion, etc) by the pathologists according to whether the follow-up examination and treatment are needed. All the malignant regions are labeled along with their contours and converted to a mask. (Shown in Figure.~\ref{fig:dataset}.)

% 所谓的malignanant【医生标了轮廓的】：在本文中指代胃黏膜中的有肿瘤倾向或者严重的癌前病变，主要包括： 
% 上皮内低级别瘤变：low grade intraepithelial neoplasia (LGIN) 
% 上皮内高级别瘤变：high grade intraepithelial neoplasia (HGIN) 
% 腺癌：adenocarcinoma (AC)  
% 腺癌中也包括了有 印戒细胞癌/低粘附性癌（少）有时候这俩会混着一起标 
% signet ring cell carcinoma (SRCC)/Poorly Cohesive Carcinoma 
% 低分化癌（也少）：poorly differentiated adenocarcinoma 
% 黏液腺癌（最少）：mucinous adenocarcinoma (MC) 
% 管状腺癌（最多）： Tubular adenocarcinoma 
% 所谓的benign【医生没标轮廓的负样本】：在本文中指代没有肿瘤倾向或病变不严重的情况，主要包括： 
% 胃炎，包括萎缩性胃炎和非萎缩性胃炎：chronic atrophic gastritis/Chronic non-atrophic gastritis/inflammation 
% 肠上皮化生：intestinal metaplasia 
% 胃息肉：gastric polyps 
% 胃黏膜糜烂：gastric mucosal erosion 
% 以上，有缩写的是上次paper里面用过的，而其他名词我没有找到【国际统一】的标准，因此可以自己写一个缩写，只对本文paper范围内使用也ok。 

\subsection{Semi-automated Annotation System}

% 图A：这张图尺寸为2233x3408 pixels
% 全手标
% 需要1~3小时

% 图B：这张图尺寸为2386x3729 pixels
% 拿普通的unet算法跑一遍需要不到20秒（绿色overlay就是）
% 然后医生在这个基础上去做修改（因为算法有错的时候），修改的时间只需要20分钟（蓝色的overlay是修正后的结果）

% 因此半自动的优点在于：
% 1，节省医生从头开始标注的时间
% 2，算法能够在一些曲了拐弯的人类难以用打点的方式标注的地方完美标注
% 3，医生在修改标注的过程，实际上也指出了算法的不足，这些算错的区域直接用作hard case
% 4，optional我瞎写的你爱用不用，建议不用，效果图是Unet没错，但是后来发现随着分割网络精度的提升，半自动标注也越来越趋近于医生手标（这不明摆的么），甚至可以批量搞出来一大堆不需要医生检查的正样本直接用来训练

The manual annotation procedures designed by the experts of Shanghai General Hospital are:
\begin{enumerate}[1)]
\item Coarse labeling. The pathologist finds the suspicious area in the high-resolution gigapixel image and selects them with bounding boxes.
\item Fine labeling. Within the selected bounding box, pathologist further confirms whether it is malignant, draws a closed curve along the contour of the malignant region and marks the lesion type of the region.
\item Double checking. Finally, the experts go through all the annotated WSIs and verify the correctness of the labels.
\end{enumerate}

\begin{figure}[h]
\centering
\includegraphics[width=1\linewidth]{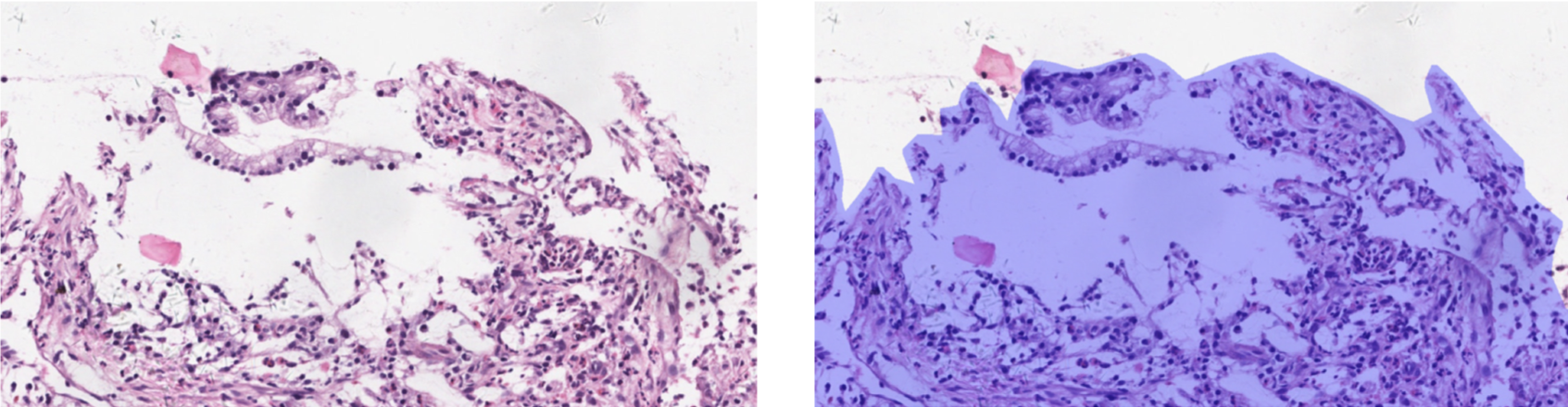}
\vspace{-1.5em}
\caption{An enlarged part of WSI. Blue overlay in the right image denotes the manual annotation from the pathologist, which may cover many empty regions.}
\vspace{-1em}
\label{fig:manul_eg}
\end{figure}

Manually labeling a WSI usually takes $1-3$ hours and the contour may not align perfectly with the border due to the enormous image size and the highly unsmooth edge (shown in Figure.~\ref{fig:manul_eg}).

% \begin{figure}[h]
% \centering
% \includegraphics[width=0.9\linewidth]{fig/semi_anno.png}
% \caption{semi-automated annotation}
% \label{fig:seg_anno_auto}
% \end{figure}

To address this issue, we introduce a semi-automated annotation system (shown in Figure.~\ref{fig:seg_anno_auto}).
An annotation refinement module with a series of image processing techniques is employed to eliminate the background areas in the manually labeled regions (shown in the right part of Figure.~\ref{fig:seg_anno_auto}). Color deconvolution technique~\cite{van2000hue,ruifrok2001quantification} is used to extract Hematoxylin \& Eosin-stained regions (image II).  Based on the output of color deconvolution, we could obtain the foreground by applying Otsu thresholding~\cite{otsu1979threshold} (image III). The final segmentation annotations are the intersection between the foregrounds and the pathologists annotated areas (image IV). These procedures could fix the empty regions and loose boundary without modifying the true malignant region.

% \begin{figure}[h]
% \centering
% \includegraphics[width=1\linewidth]{fig/refine.png}
% \caption{segmentation ground-truth refinement}
% \label{fig:seg_anno_refine}
% \end{figure}

As shown in the left part of Figure.~\ref{fig:seg_anno_auto}, we further propose to initialize a preliminary annotation with the output of a segmentation network trained on a few labeled WSIs. Then, pathologists only need to modify the masked regions discovered by the segmentation network, which could effectively speed up the labeling procedure.
Basically, our semi-automated annotation system could generate more accurate region contours and reduce the labeling time to  $~20$ minutes.

With the protocol and semi-automated tools designed for acquiring data, we collected $518$ WSIs with both slide-level labels and detailed region annotation from Shanghai General Hospital in the year of 2018. Among these WSIs, $207$ are labeled positive, and $311$ are labeled negative.

\subsection{Large-scale Dataset from Multiple Institutes}
Moreover, in order to test the generalizability of our proposed model in real-world scenario, we collected $10,315$ WSIs with slide-level screening labels from four hospitals in East China, i.e., Sijing Hospital of Shanghai Songjiang District (SHSSD), Songjiang District Center Hospital (SDCH), Zhejiang Provincial People’s Hospital (ZPPH) and Shanghai General Hospital (SGH). These hospitals have their own pathology slide preparation procedures, which make the WSIs have different tissue thicknesses and color tones.

%------------------------------------------------------------------------

\begin{figure*}[thb]
\centering
\includegraphics[width=1\linewidth]{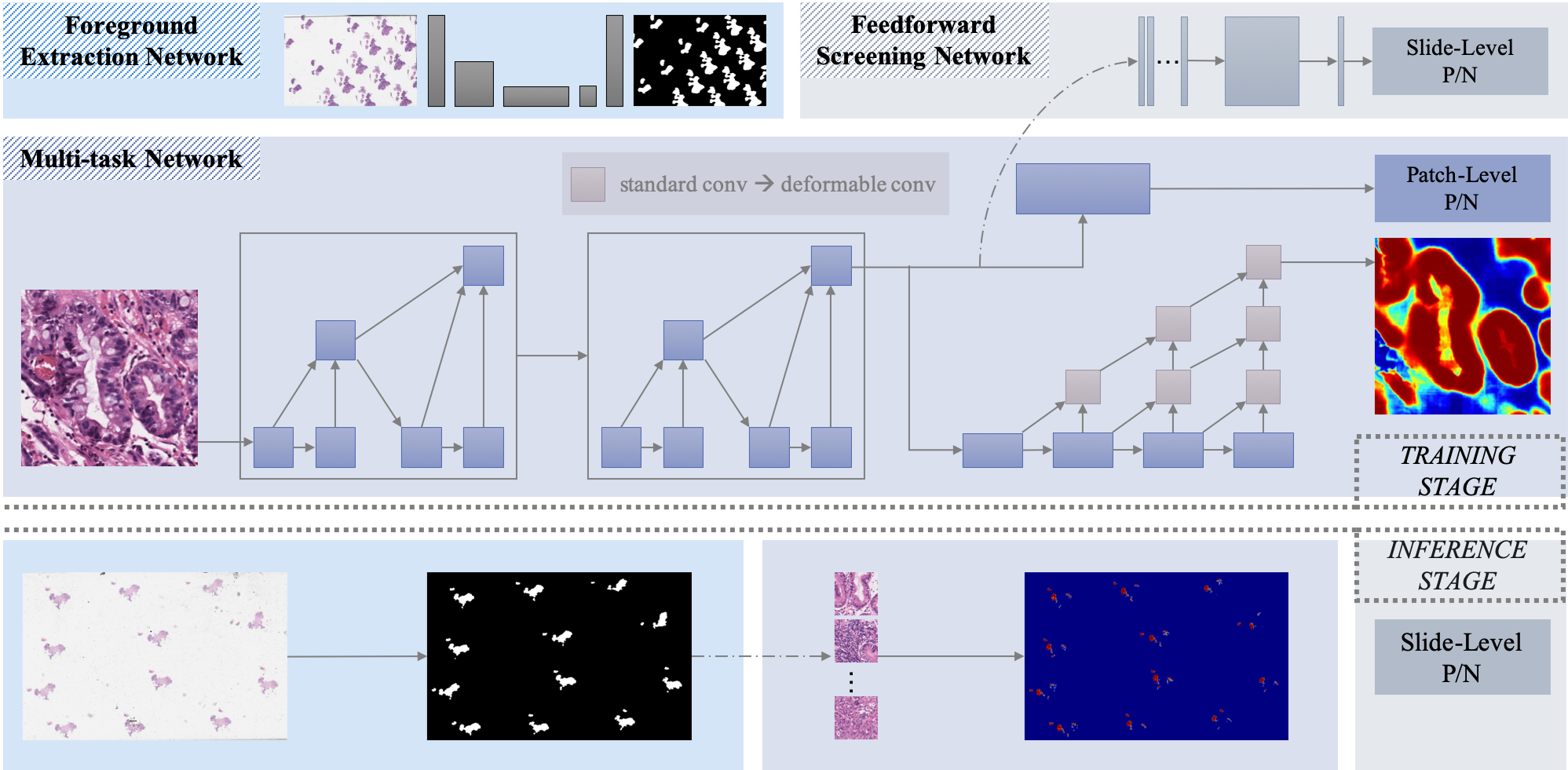}
\caption{The overview of our proposed gastric cancer screening framework. On the top, we show the structure of the proposed foreground extraction network, multi-task network, and feedforward screening network, which are trained separately during the training stage. The inference stage is illustrated in the bottom. The mask is obtained by the foreground extraction network. Then the patches cropped from the foreground are send to the multi-task network for classification and segmentation. Finally, the feature vectors from the second network are collected and fed to the screening network to produce the slide-level result.}
\label{fig:overview}
\end{figure*}

\section{Methods}

The main challenge in analyzing WSIs is that the images are extremely large, while only a small part of images contains the lesion region. Hence, we propose a gastric cancer screening and localization framework (shown in Figure.~\ref{fig:overview}), which consists of 1) a lite segmentation network for extracting the tissue region and eliminating the background area, 2) a multi-task network for generating classification and segmentation results for every cropped patch, and 3) a simple feedforward network for providing the slide-level screening result.

\subsection{Network Structure}

%训练数据来源：提取大图中的每个patch的特征向量（分割网络分类头中介于全连接层和平均池化层之间的feature_map）及置信度，作为小分类网络的输入。

\textbf{Multi-task Network}. We employ deep layer aggregation (DLA) structure~\cite{yu2018deep} as the backbone since it is designed to aggregate layers to fuse semantic and spatial information for better recognition and localization. 
In our case, we utilize the dense prediction network for lesion region segmentation and change it to a multi-task structure by adding a classification branch on the output of the encoder.
In the training phase, we use the patches cropped from ROIs with their annotated malignant region masks as the positive inputs and the patches randomly cropped from the benign region with the background (all-zero) masks as the negative inputs, and train the network in an end-to-end manner by combining the classification loss and the segmentation loss (Equation.~\ref{eq:loss}). 
In the inference phase, patches from each sliding window are sent to the network to generate the classification score and the segmentation heatmap.

\begin{equation}
L = \lambda L_{cls} + L_{seg}
\label{eq:loss}
\end{equation}
where $L_{cls}$ denotes the loss in classification branch, $L_{seg}$ is the binary cross entropy loss in segmentation branch, and $\lambda$ is a hyper-parameter to balance the weights of these two losses.

\textbf{Feedforward Screening Network}. The input of this network is a matrix $\mathbf{X}\in \mathbb{R}^{k\times d}$ which consists of the mid-layer features (i.e., the output of encoder) $\mathbf{x}\in \mathbb{R}^{1\times d}$ in the dense prediction DLA structure from $k$ patches with the highest probability of being malignant. 
The network contains a $1\times1$ convolution layer with a ReLU layer, a max-pooling operation along the sample axis which generates a $\mathbf{Y}\in \mathbb{R}^{1\times d}$ tensor, and a fully connected layer followed by a sigmoid function. 
Ground truth label of this network is defined in a multi-instance learning (MIL) manner, i.e., whether the WSI contains at least one malignant region. 
% Input in the training phase consists of the feature vectors from $k$ patches with the highest probability of being malignant. While in the inference phase, we pick $\frac{k}{2}$ high probability patches and choose another $\frac{k}{2}$ randomly for stability.
 
Since lesion regions only account for $20\%$ percent of the tissue or less,  it is difficult for a network targeting at recognition task such as as~\cite{krizhevsky2012imagenet,he2016deep,szegedy2016rethinking} to find where to pay attention. In this project, we could take advantage of the detailed lesion region annotation. The multi-task network is designed for producing patch-level results. The segmentation task could not only generate the lesion region mask but also help the whole network locate the regions that really need to focus, which could implicitly support the classification branch to achieve higher performance.  The feedforward screening network is proposed for generating the slide-level results. The basic idea in designing this simple feedforward network follows the concept of MIL. 
%Furthermore, the candidate selection scheme in the inference phase is a consequence of considering the robustness.

Because the gastric WSIs usually carry many blank areas. We further design a \textbf{foreground extraction network} to extract all tissue region and reduce the running time in the inference phase. It is a segmentation network with $4$ convolution layers and $1$ upsampling layer. 
The ground truth masks are firstly generated by color deconvlotion~\cite{van2000hue,ruifrok2001quantification} and Otsu thresholding~\cite{otsu1979threshold}. Then we manually go through all the algorithm generated masks and delete failure cases caused by some unusual problems such as shadows on the edge of the slides.
The input of this network are the resized WSIs from the lowest magnification level (i.e., $\times 0.635$). The output of this network is utilized in the inference stage for selecting the sliding window area.

\subsection{Deformable Convolution}

Another issue in WSIs is that the lesion regions often come with irregular shapes and various sizes while standard convolution operation always fails to handle large geometric transformations due to the fixed geometric structures. Therefore, instead of using standard convolution operation, we employ the deformable convolution layer~\cite{dai2017deformable} in the decoder.

\begin{figure}[h]
\centering
\includegraphics[width=0.95\linewidth]{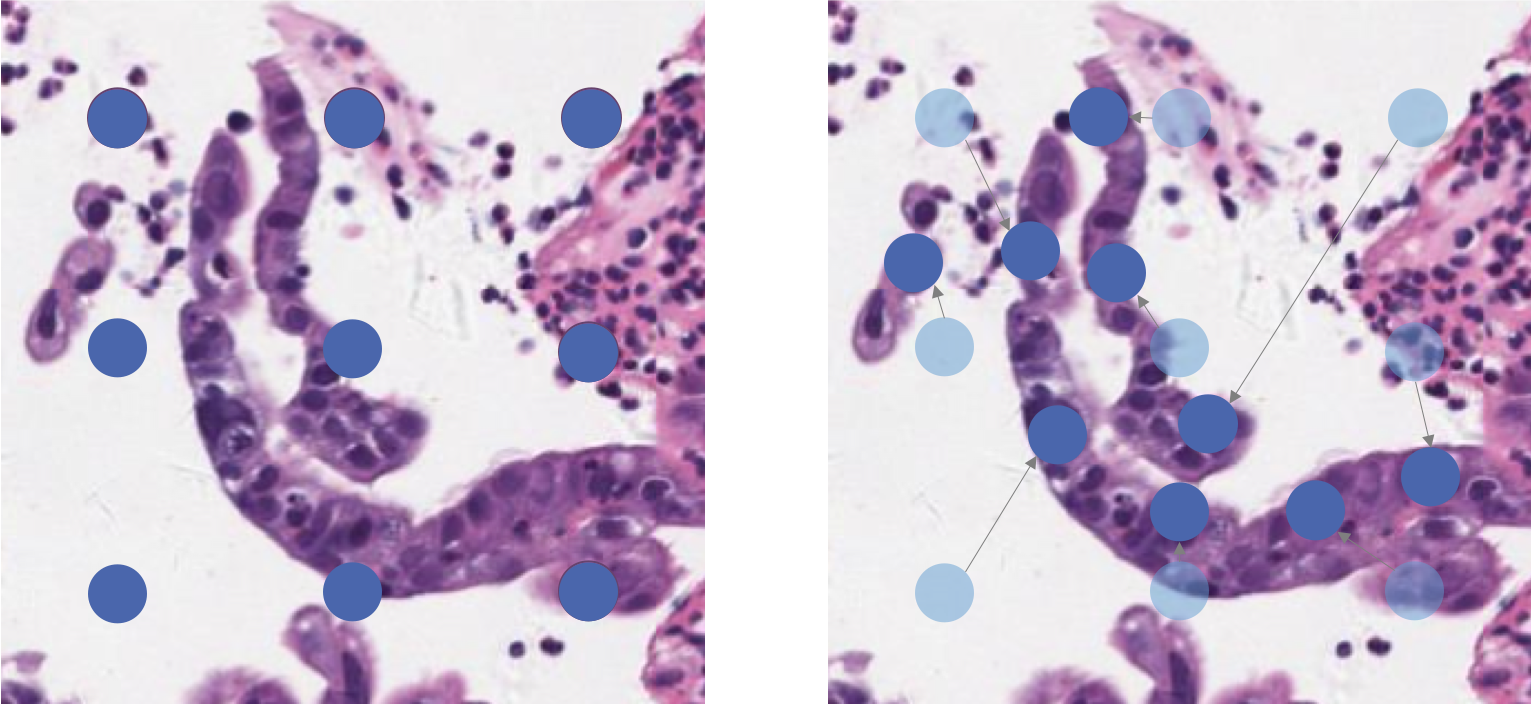}
\caption{Left: standard convolution operator. Right: deformable convolution operator.}
\label{fig:defo_conv}
\end{figure}

Deformable convolution is proposed based on the idea of augmenting the spatial sampling locations in the modules with additional offsets:

\begin{equation}
y(p_0) = w(p_n) \cdot x(p_0 + p_n + \Delta p_n)
\end{equation}
where  $x$ denotes the input feature map, $y$ denotes the output feature map, $p_0$ is a location on input feature map, $p_n \in \mathcal{R}$ and $\mathcal{R}$ is a regular grid $\mathcal{R} = {(-1,-1), (-1,0), \dots, (0,1), (1,1)}$.
This type of convolution operation allows adding 2D offsets $\Delta p_n$ to the regular grid sampling locations $p_0 + p_n$ in the standard convolution. 

As demonstrated in Figure.~\ref{fig:defo_conv}, we could easily benefit from the adaptive receptive field when handling the lesion regions with irregular shapes. The deformable convolution operation enables free form deformation of the sampling grid by adding additional offsets (the arrows in the right image of Figure.~\ref{fig:defo_conv}). The offsets are learned from the segmentation task simultaneously with convolutional kernels during training. Hence, the deformation is conditioned on the input features in a local, dense, and adaptive way.

As a consequence,  we could capture the feature of the tissue regions with various shapes by utilizing deformable convolution operation, which leads to better performance on the segmentation task.

% \subsection{Patch Concatenation in Dense Prediction}

% \begin{figure}[h]
% \centering
% \includegraphics[width=0.8\linewidth]{fig/patch_problem.png}
% \caption{sudden change}
% \label{fig:patch_problem}
% \end{figure}

% Post-processing is a crucial part of WSI segmentation task. Selecting patches without overlap usually lead to a severe problem, the heatmaps of segmentation discontinue at the edges of each patch (shown in Figure.~\ref{fig:patch_problem}).

% To solve this problem, we design a simple but effective trick in the sliding window process (shown Figure.~\ref{fig:patch_solution}).

% \begin{figure}[h]
% \centering
% \includegraphics[width=0.5\linewidth]{fig/patch_trick.png}
% \caption{patch connection}
% \label{fig:patch_solution}
% \end{figure}

% \begin{itemize}
% \item Patches are designed to have overlap with each other in order to ensure there is enough shared information between the neighbor patches. As a result, the overlapped area has information from both side and tend to be smooth and continuous.
% \item After generating these segmentation heatmaps, without any additional computation, we use the heatmap of the left patch for the left half overlapping region and the heatmap of the right patch for the right half overlapping region.
% \end{itemize}

% Although this design increases the time complexity from $O(\frac{\mathcal{S}}{w})$ to $O(\frac{\mathcal{S}}{s})$, it avoids any merging/averaging operation but utilizing the direct output from the network.

%------------------------------------------------------------------------

\section{Evaluation}

\subsection{Implementation Details}
We choose the basic DLA-34 with dense prediction part as the backbone of the multi-task network, replace the first $7\times7$ convolution layer with three $3\times3$ convolution layers, and several standard convolution layers in decoder with $3\times3$ deformable convolution layers. Besides, in-place activated batch normalization~\cite{rota2018place} is employed to reduce the memory requirement in training deep networks. 

All WSIs are resized to $512 \times 512$ for foreground extraction. The input patch size of multi-task network is also set to $512 \times 512$. The output of the classification branch of this network is a $512$ feature vector. 

The $\lambda$ in Equation.~\ref{eq:loss} used for balancing the classification loss and segmentation loss is set to $0.01$.
The hyper-parameters in the feedforward screening network are selected by cross-validation. We pick $k=64$ to form a $64\times512$ matrix as the input of the feedforward screening network. 
Threshold of feedforward screening network is set to $0.74$, i.e., the WSI should be considered as positive if the output of the sigmoid function is larger than $0.74$.

We use Adam optimizer to train the networks separately.  The initial learning rate is set to $10^{-3}$ and reduced by a factor $0.5$ when the validation loss stagnates for $3$ epochs. 
The foreground extraction network is trained for $100$ epochs with the batch size $32$ on one GPU. The multi-task network is trained for $100$ epochs with the batch size $16$ on $4$ GPUs with $12$GB RAM. And the feedforward screening network is trained for $20$ epochs with the batch size $32$ on a single GPU.

% 重新核实，
% '''
% 分割
% Training:
% wsi数量： 阳性：146 阴性：218    病人对应roi的数量： 阳性：299 阴性：2724        切分roi的patch数量： 阳性：26140 阴性：127787
% Validation：
% wsi数量： 阳性：61 阴性：93    病人对应roi的数量： 阳性：101 阴性：1167        切分roi的patch数量： 阳性：5367 阴性：40698
% '''
% ‘’‘
% 筛查（分类）
% Training：
% wsi数量：阳性：146 阴性：218
% Validation：
% wsi数量： 阳性：61 阴性：93
% Testing：
% wsi数量： 阳性：102 阴性：838     这部分需要再仔细统计一下 目测小规模筛查测试数据集900+  大规模筛查测试集需要和迪核对
% 大规模： 阳性：268 阴性：4831 
% ’‘’

\subsection{Evaluation of Segmentation}

% \textbf{Experiment Settings.}
$21,507$ positive patches are cropped from the annotated $400$ ROIs, and $168,485$ negative patches are collected from the WSIs that are diagnosed as benign.  Data augmentation techniques such as horizontal/vertical flip, rotation, random cropping and resizing, changing the aspect ratio and image contrast, and adding Gaussian noise are applied during training.
We use repeated random sub-sampling strategy in evaluation. Patches from about $30\%$ WSIs are selected as the testing set by the patient ID, and the whole procedures are repeated $4$ times for stable results. 

For the evaluation metric, we use the Dice coefficient (DSC) defined over image pixels.

\begin{equation}
\displaystyle DSC=\frac{2TP}{2TP+FP+FN}
\end{equation}

We compare our proposed method with commonly used segmentation networks, i.e., FCN~\cite{long2015fully}, U-Net~\cite{ronneberger2015u}, and the standard DLA~\cite{yu2018deep} structure without any modification.

% \textit{!!!!change to a pyplot bar figure!!!!!}
% \begin{table}[h]
% \centering
% \begin{tabular}{l r}
% \hline
% % \multicolumn{2}{c}{Placeholder} \\
% % \cline{1-2}
%     & DSC \\
% \hline
% FCN        & 0.8109     \\
% U-Net       & 0.7847     \\
% DLA        & 0.8248     \\
% Ours  & 0.8331     \\
% \hline
% \end{tabular}
% \caption{segmentation results of our proposed model and other comparison methods.}
% \label{tbl:seg_res}
% \end{table}

\begin{figure}[h]
\centering
\includegraphics[width=0.85\linewidth]{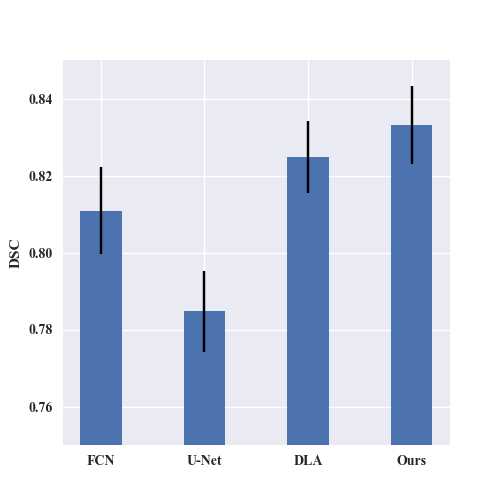}
\caption{segmentation results of our proposed model and other comparison methods.}
\label{fig:seg_res}
\end{figure}

FCN and U-Net serve as the baselines, which give us a sense about the performance of commonly used methods on our dataset. Dice coefficient of these two methods is $0.8109\pm0.0113$ and $0.7847\pm0.0106$, respectively.
The possible reason for FCN performing better could be it benefits more from classification branch without the skip connections like U-Net.
DLA performs better, i.e., $0.8248\pm0.0094$, because it benefits from fusing information by aggregating layers. 
Our proposed model achieves the highest DSC of $0.8331\pm0.0101$ as we adjust the standard DLA structure according to our problem settings by replacing the standard convolution with deformable convolution operator.

% XXX need a higher res image inf pdf format
\begin{figure*}[h]
\centering
\includegraphics[width=1\linewidth]{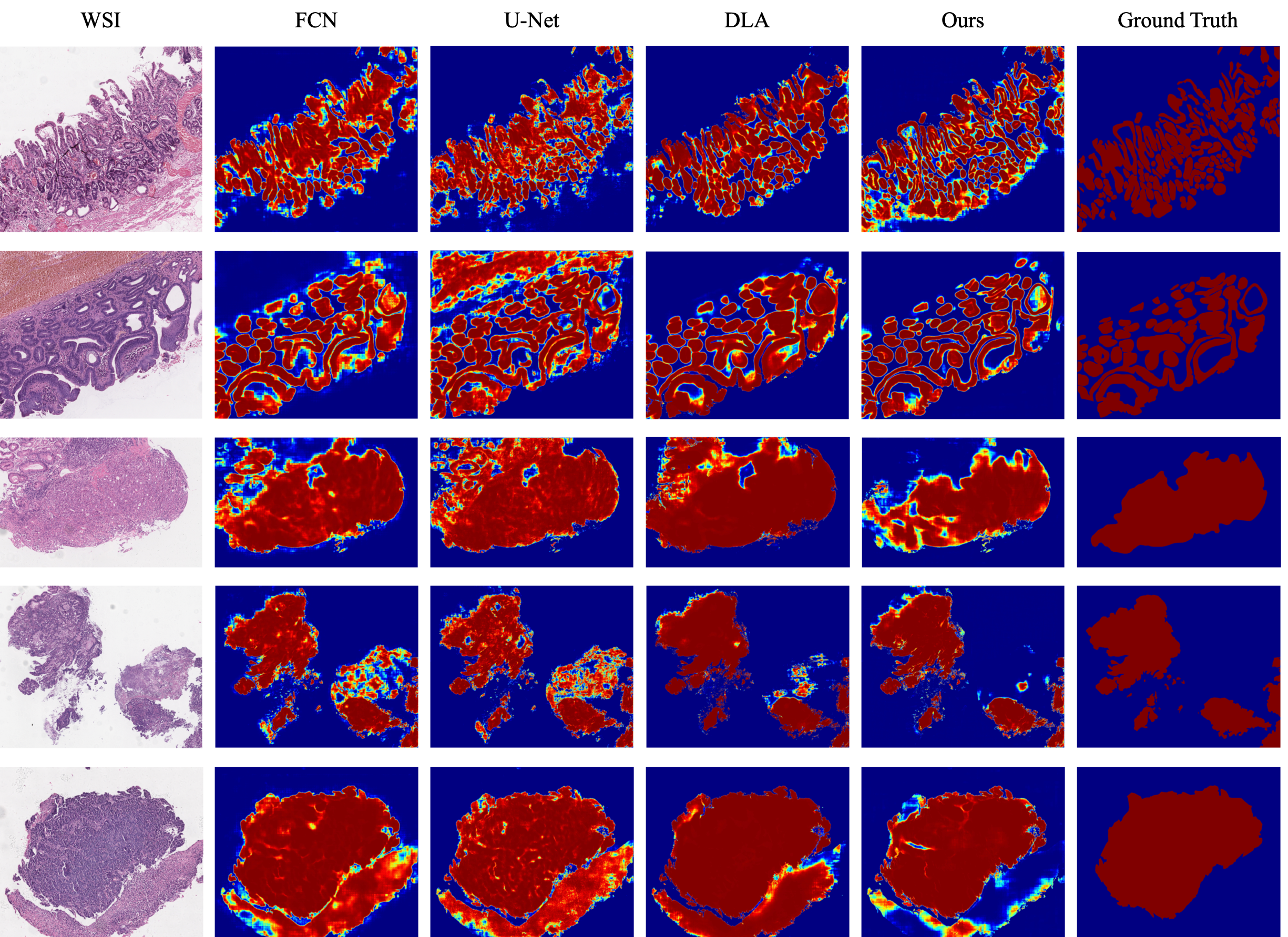}
\caption{Examples of segmentation results. We show $5$ ROIs with their original image (left) and the ground truth masks (right) labeled by pathologists. Results of our proposed method and other comparison methods are illustrated in the middle columns.}
\label{fig:seg_demo}
\end{figure*}

To better demonstrate the effectiveness of our proposed method, we show some examples in Figure.~\ref{fig:seg_demo}. FCN and U-Net often generate unclear and unsmooth boundaries (1st and 2nd row). As we could see from the 3rd to the 5th column, the 3 comparison methods make a lot of false positives compared to our method. In the 2nd row, U-Net even includes red blood cell area in the result. In the 5th row, our model is the only one that successfully identifies the tissue on the bottom to be negative.

\subsection{Evaluation of Screening}

% \textbf{Experiment Settings.}
Compare with the segmentation evaluation that requires a testing set with lesion region annotation, evaluating the screening results of our proposed model could be done on a larger testing set with screening annotation only, i.e., $102$ positive samples and $838$ negative samples.

In Table.~\ref{tbl:cls_res_small}, we report the screening results of our proposed framework and comparison methods on multiple evaluation metrics, i.e., sensitivity and specificity at the threshold of $0.74$, and Area Under Precision-Recall Curve (AUC). 
The comparison methods are 1) single task classification networks, such as ResNet~\cite{he2016deep}, DenseNet~\cite{huang2017densely}, Res2Net~\cite{gao2019res2net}, EfficientNet~\cite{tan2019efficientnet}, and DLA (the same backbone as our proposed network structure); 2) multi-task network structure with different backbones, such as FCN~\cite{long2015fully} and U-Net~\cite{ronneberger2015u}.
\begin{table}[h]
\centering
\begin{tabular}{l r r r r}
\hline
% \multicolumn{2}{c}{Placeholder} \\
% \cline{1-2}
     & Sensitivity & Specificity & AUC \\
\hline
ResNet-34   & 0.9118 & 0.9427 & 0.9017 \\
DenseNet-121    & 0.9118 & 0.9463 & 0.9021 \\
Res2Net-50  & 0.9314 & 0.9606 & 0.9047 \\
EfficientNet-B4 & 0.9509 & \textbf{0.9653} & 0.9081 \\
DLA       & 0.9314  & 0.9582 & 0.9038    \\
\hline
FCN        & 0.9608  & 0.8938  & 0.9099   \\
U-Net      & 0.9608  & 0.8914  & 0.9117    \\
Ours    & \textbf{0.9705}   & 0.9272  & \textbf{0.9166}    \\
\hline
\end{tabular}
\caption{Screening results of our proposed model compared with other methods.}
\label{tbl:cls_res_small}
\end{table}

The primary goal of our proposed framework is gastric cancer screening. In other words, the number of negative samples is much larger than the positive ones (about $9:1$ in this testing set).
In this case, sensitivity is the most critical evaluation metric when we take clinical background knowledge into consideration since we would not like to miss any of the malignant cases. Besides,  specificity is the second priority because we do not want too many false alarms, either. AUC is a metric that expresses both sensitivity and specificity.
Although the single task classification networks like EfficientNet-B4 could achieve the highest specificity of $96.53\%$, it is not a suitable choice as it generates too many false negative.

For the multi-task networks with different backbones, DLA structure only has $19.4$ million parameters. FCN and U-Net have $24$ million and $31$ million, respectively. Adding the classification branch only brings about $2,000$ more parameters. So the final model size is less than $20$ million parameters.

The proposed method shows the highest sensitivity of $97.05\% (95\% CI, 91.02\%-99.24\%)$, the second-best specificity of $92.72\% (95\% CI,90.69\%-94.34\%)$, and the leading AUC of $0.9872$, which outperforms others methods by a large margin.

\subsection{Evaluation in Real-world Scenario}

Also, we test our best model on our large-scale real-world set collected from $4$ medical centers.
Table.~\ref{tbl:hsp_num} shows the numbers of images of the collected data. 

\begin{table}[h]
\centering
\begin{tabular}{c c | r r r}
\hline

  & Year & Total & Positive & Negative  \\
\hline
\multirow{3}{2.5em}{\textbf{SGH}}  & 2015 & 2083      & 85  & 1998     \\
& 2018       & 3207      & 299  & 2908     \\
& 2019       & 3670      & 50  & 3620     \\
\multicolumn{2}{c|}{\textbf{SHSSD}}      & 495      & 8  & 487     \\
\multicolumn{2}{c|}{\textbf{SDCH}}       & 391      & 6  & 385     \\
\multicolumn{2}{c|}{\textbf{ZPPH}}       & 469      & 8  & 461     \\
\hline
\end{tabular}
\caption{Number of WSIs collected from the four institutes in different years.}
\label{tbl:hsp_num}
\end{table}

All the training images with lesion region annotation are from SGH in the year 2018. Besides those training images, we further collect $3,207$ images in that year, $3,670$ images in 2019 as the most recent samples, and $2,083$ images in 2015 as the old samples since they did not use too many automated devices for fixation, sectioning, and staining at that time. Moreover, to test the generalization ability of our proposed model, we collect $1,356$ images from $3$ other hospitals, i.e., SHSSD, SDCH, and ZPPH. The devices and procedures in making histology slides are different in these hospitals which may affect the final WSIs.
Overall, we have $10,315$ WSIs from $4$ hospitals and years, and the positive ratio is less than $5\%$. 

\begin{figure}[h]
\centering
\includegraphics[width=1\linewidth]{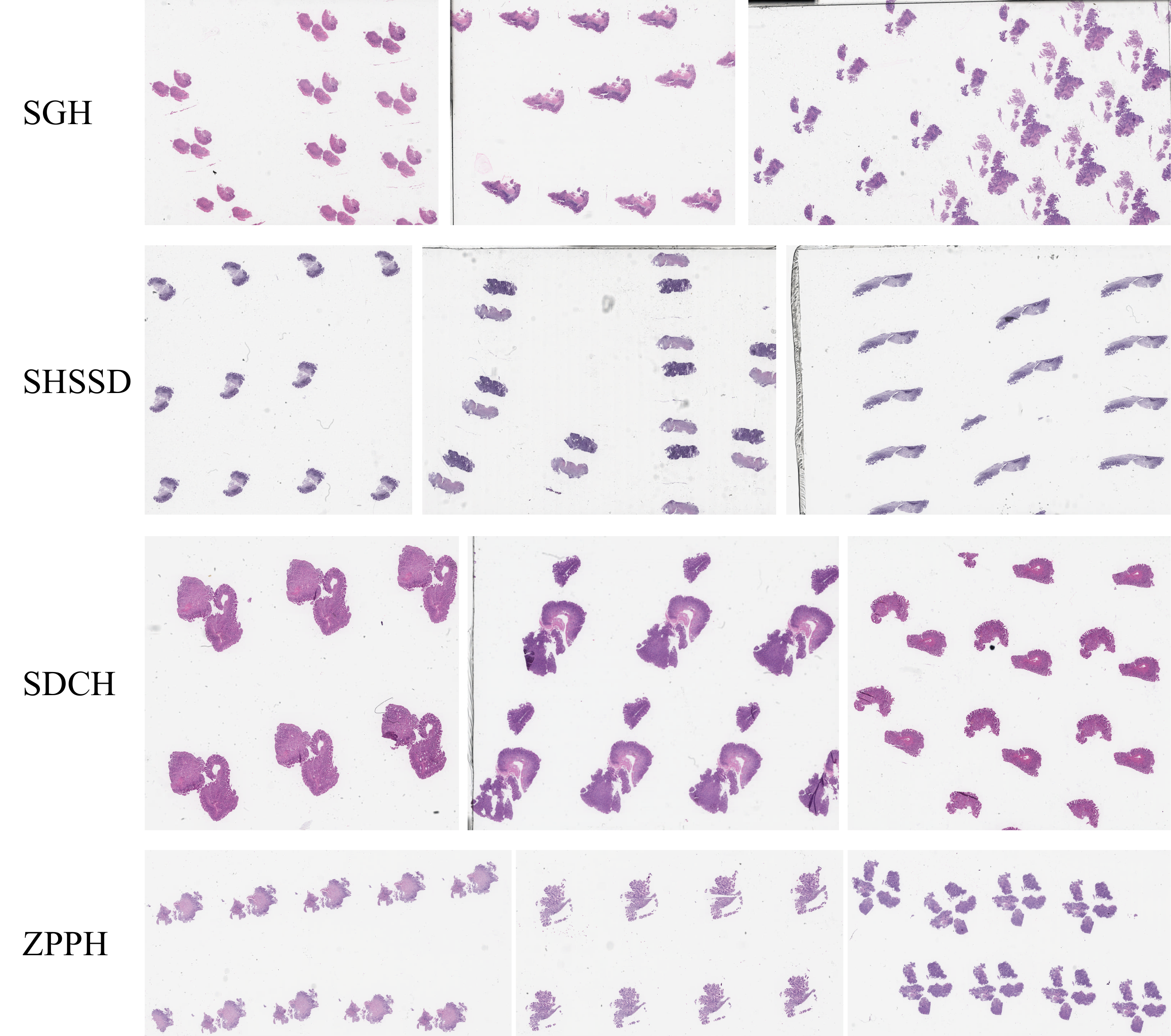}
\caption{WSI examples from four institutes.}
\label{fig:multi_center}
\end{figure}

\begin{table}[h]
\centering
\begin{tabular}{c c | r r r}
\hline
% \multicolumn{2}{c}{Placeholder} \\
% \cline{1-2}
    & Year & Sensitivity & Specificity  \\
\hline
\multirow{3}{2.5em}{\textbf{SGH}}   & 2015 & 1.0000  & 0.7753     \\
                            & 2018       & 0.9967  & 0.7995   \\
                            & 2019       & 1.0000  & 0.7829  \\
\multicolumn{2}{c|}{\textbf{SHSSD}}      & 1.0000  & 0.7371    \\
\multicolumn{2}{c|}{\textbf{SDCH}}       & 1.0000  & 0.6571    \\
\multicolumn{2}{c|}{\textbf{ZPPH}}       & 1.0000  & 0.7527     \\
\hline
\end{tabular}
\caption{Screening results in real-world scenario.}
\label{tbl:cls_res_real}
\end{table}

We apply our best model on these data and present sensitivity, and specificity in Table.~\ref{tbl:cls_res_real}. 
We achieve the sensitivity of $100.00\%$ in $5$ testing set, $99.67\%$ in the other, and the specificity is around $75\%$ in most of the cases. 

However, the data distribution in the real-world scenario is different from our training and validation dataset, i.e., less positive samples and more outliers such as out of focus samples. 

Moreover, the procedures in slide preparation also cause the differences (shown in Figure.~\ref{fig:multi_center}). The sectioned tissues of SDCH are thicker than others, and they are stained much darker, which lead to the lowest specificity. WSIs from SHSSD are more blueish than SGH. Besides SGH, ZPPH also uses the machines for automatic sectioning and staining, so the performance on ZPPH data is higher than the other two institutes.

%------------------------------------------------------------------------
\section{Discussion of Failure Cases}

\subsection{Out of Focus Slides}

\begin{figure}[h]
\centering
\includegraphics[width=0.9\linewidth]{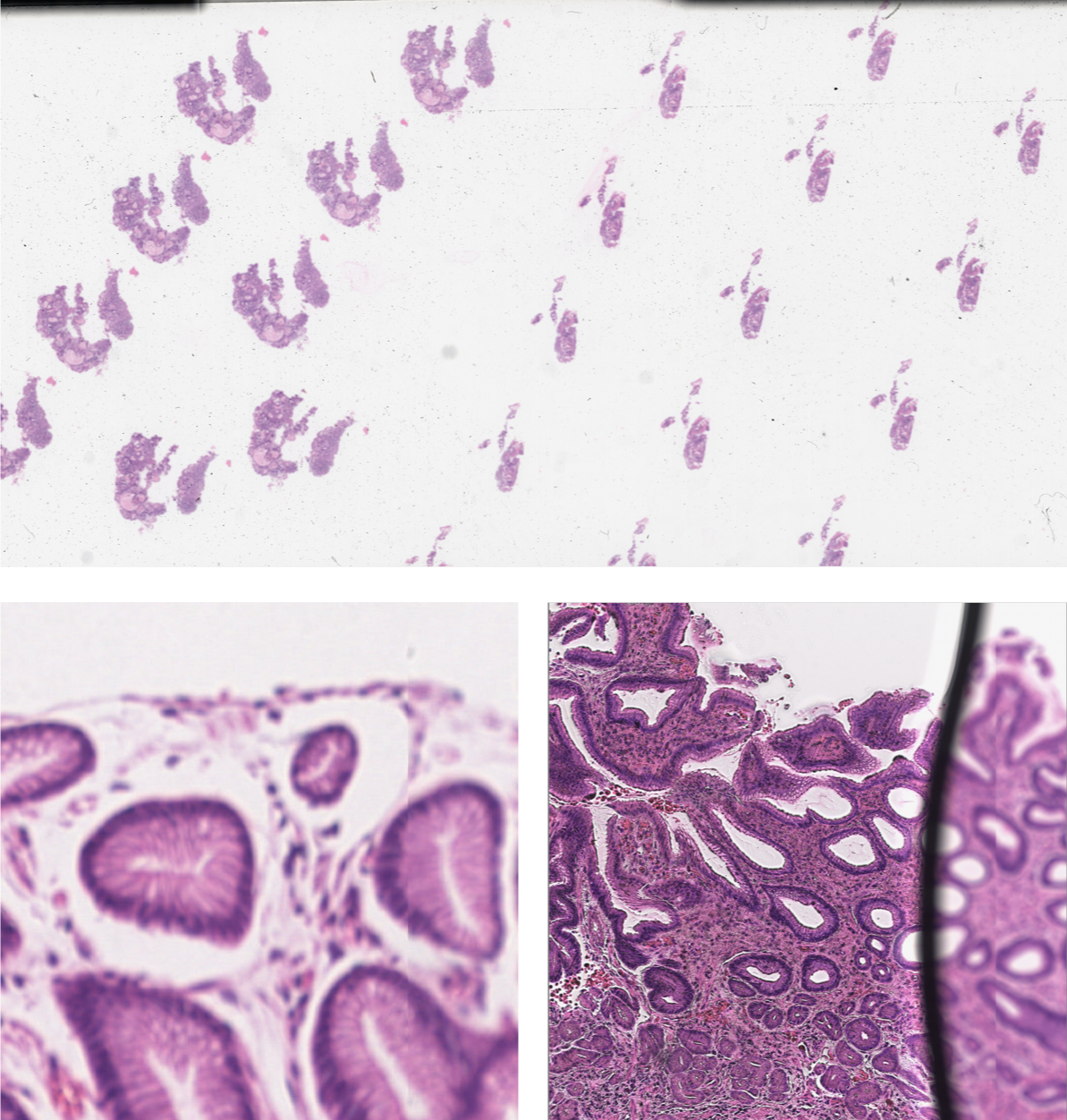}
\caption{Top: an out of focus WSI. Left: a patch of the blurred image.  Right: partially out of focus example.}
\label{fig:outfocus}
\end{figure}

As shown in Figure.~\ref{fig:outfocus}, Our proposed model fails on these blurred images. The blurry is caused by out of focus during scanning. Instead of focusing on the tissue sections, the scanner focuses on the dust. 

There is no effective way to classify the out of focus images correctly since they are highly blurred. So we add a blurry detection module, i.e., if it is a blurred image, there is no need to continue to classify whether it is positive or negative.
We manually pick $1324$ blurred images (totally blurred and partially blurred) and $5512$ clear images to train a small neural network with $2$ $5 \times 5$ convolution layers and a fully connected layer.

\subsection{Failure Cases by Subtypes}

Collaborating pathologists analyzed our testing results and found that errors are more from certain types of disease.

\begin{figure}[h]
\centering
\includegraphics[width=0.9\linewidth]{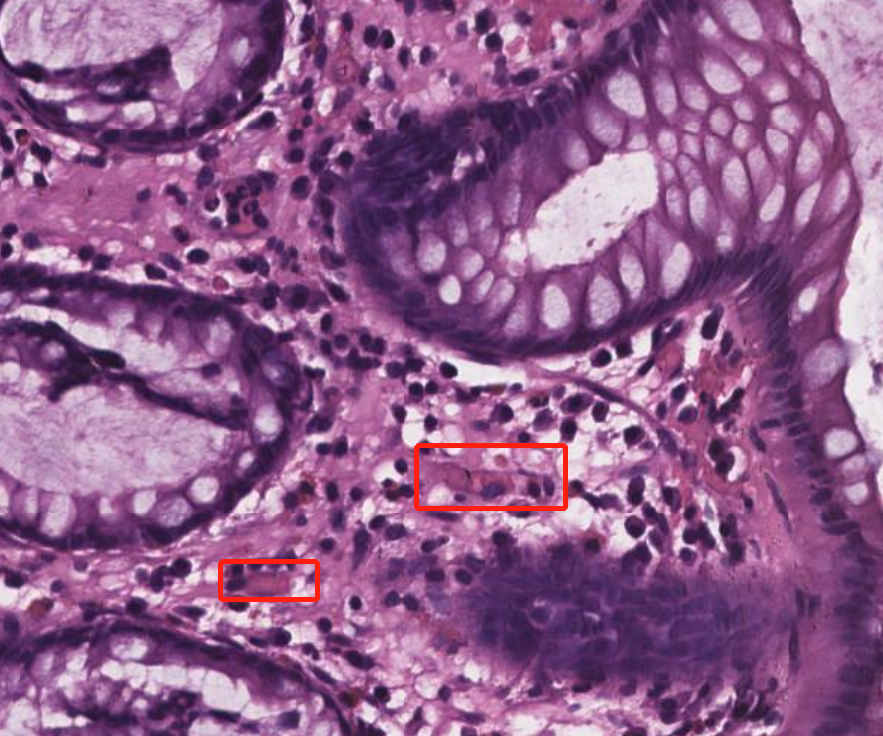}
\caption{Signet ring cell carcinoma (signet ring cell in red bounding box).}
\label{fig:fn}
\end{figure}

Samples from signet ring cell carcinoma are easily missed since there are not many signet ring cells in the WSIs, and the size of cells is too small compare with the whole tissue. Even pathologists need further examination to diagnose (shown in Figure.~\ref{fig:fn}). 

\begin{figure}[h]
\centering
\includegraphics[width=0.9\linewidth]{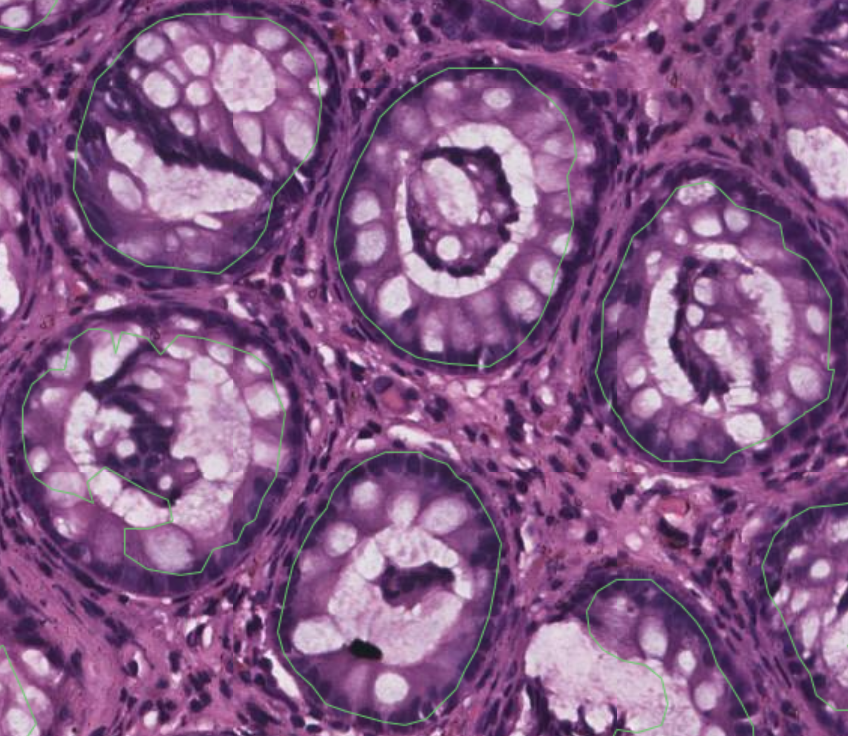}
\caption{Intestinal metaplasia. Regions in the green contours are identified as cancer by the algorithm.}
\label{fig:fp}
\end{figure}

Some of intestinal metaplasia cases are classified as positive because intestinal metaplasia is a precancerous condition, i.e., ``borderline'' between positive and negative (shown in Figure.~\ref{fig:fp}). 

To solve this issue, we actively communicate with pathologists to get more of these types of samples to add to the training set and try to improve the performance of our algorithm on these subtypes.

%------------------------------------------------------------------------

\section{Conclusion and Future Work}
In conclusion, with the assumption that detailed lesion region annotation is necessary for the network to locate the crucial part in the gigapixel resolution WSIs and achieve the better performance, we designed a semi-automated annotation system and collected a large dataset consisting of $518$ WSIs with lesion region masks and over $10,000$ samples with screening results. To exploit our dataset, we propose a gastric slides screening framework to reduce the workload of pathologists and lower the inter-observer variability. The whole framework consists of three networks, a multi-task network for patch-level classification and segmentation, a 3-layer feed-forward network for slide-level screening result, and a simple segmentation network for foreground extraction. 

One of the possible future direction of this project is cancer subtype diagnosis. Instead of just providing the ``positive/negative'' screening result, an automated lesion type diagnosis system could present the result like ``adenocarcinoma'', ``signet ring cell carcinoma'', ``chronic atrophic gastritis'', etc.  Furthermore, since a single WSI may contain multiple lesion types, displaying the labels on each segmented region/ROI could be more helpful, which requires an instance-level segmentation structure.

\bibliography{mybibfile}

\end{document}